\documentclass{elsart}
\input epsf
\def\gfe2{\gamma -  Fe_2 O_3}
\journal{Journal of Magnetism and Magnetic Materials}

\begin{document}
\begin{frontmatter}
\title{Magnetic properties of polypyrrole - coated iron oxide nanoparticles}
\author{Raksha Sharma,} 
\author{Subhalakshmi Lamba and}
\author{ S.  Annapoorni}
\address{Department of Physics and Astrophysics, University of Delhi, Delhi, India:110007}

\begin{abstract}
{Iron oxide nanoparticles were  prepared by sol - gel process. In-situ polymerization of pyrrole monomer in the presence of oxygen in iron oxide - ethanol suspension resulted in a iron-oxide polypyrrole nanocomposite. The structure and magnetic properties of the  nanocomposites with varying pyrrole concentrations are investigated. The X-Ray diffraction studies indicate the presence of $\gfe2$ phase  for the concentrations investigated. FTIR studies confirm the presence of polypyrrole.  The TEM studies show  agglomeration in uncoated samples and in  samples with a lower concentration of polypyrrole. Agglomeration is much reduced for samples coated with higher concentration of  polypyrrole.  The ac susceptibility measurements performed in the temperature range  $77 $ - $300$ K shows the presence of blocking,  indicating the superparamagnetic phase. The blocking temperature is found to depend on the pyrrole concentration. Monte Carlo studies for an array of polydispersed single domain magnetic particles, based  on an interacting random anisotropy model were  also carried out  and the blocking temperatures obtained from the  simulation of the ZFC-FC magnetization compares  favorably with experimental results.}
\end{abstract}


\end{frontmatter}

\section{Introduction}
Over the recent years fine particle magnetic systems with particle sizes in the range of $5$ to $50$ nm  have generated a lot of interest because of wide ranging technological applications like magnetic recording, electromagnetic shielding, sensors and magnetic refrigeration \cite{d1,s1}. They exhibit a wide range of magnetic phases like ferromagnetism, antiferromagnetism, spin glasses etc. as well as the typical superparamagnetic behavior associated with single domain magnetic particles. The properties of these systems are sensitive to  the particle size, interparticle interactions  and temperature.   
The most commonly studied magnetic nanomaterials are the oxides, in the form of ferrites and substituted ferrites. They are prepared by various processes like hydrothermal precipitation, sol - gel, sonochemical methods, sputtering, electron beam evaporation etc \cite{f1,f2,p1,p2}. Among the  standard methods,  the sol - gel method has the advantage of good composition control and a low processing temperature. 
A critical obstacle in assembling and maintaining a nanoscale magnetic material is usually its tendency to agglomerate, which is a deterrent to its application for magnetic storage. This obstacle can be overcome if the particles are dispersed in  a polymer matrix. Such nanocomposite materials are  useful  because of the processable nature of the  polymers,
  a lesser  tendency to agglomeration  and  a  uniform particle size distribution in comparison to  conventional nanomaterials. 

Iron oxide polpyrrole nanocomposites can be prepared by several methods. In our laboratory these composites have been prepared by simultaneous gelation and polymerization\cite{k1}. The magnetic phase obtained was found to be  very sensitive to the pyrrole concentration. 
 We observed that these nanocomposites showed a change of phase to the non-magnetic $\alpha -  Fe_2 O_3$ phase for certain pyrrole concentrations. However since we wish to preserve the magnetic phase, we have now prepared the composites by allowing the polymerisation to occur with  oxygen as the oxidizing agent, without using a soluble oxidant.  Earlier workers have studied the properties of metal oxides like   $CuO$, $CeO_2$ ,$NiO$, $SiO_2$ and $\alpha - Fe_2 O_3$ coated with PPy without soluble oxidants  \cite{k3,k4}.  In the case of iron oxide particles it is expected that the polymer coating will  result in  well separated, smaller sized superparamagnetic particles which can be a good candidate for magnetic refrigeration and  electromagnetic shielding  \cite{k5}.

Typically, transition metal oxide nanoparticles are characterized by a uniaxial anisotropy, and the relaxation between the two easy directions of magnetization leads to superparamagnetic behavior which is characterized by the blocking temperature. Theoretical research and modeling  in nanomagnetic materials is somewhat restricted by the competing effects of disorder (in the shape, size and position of the particles), anisotropy    and interactions, like the  dipolar interactions and the exchange interactions. The systems have been investigated  using mean field methods and thermodynamic perturbation theory \cite{newp1,newp2,newp3}. In the light of the complexity of the system, numerical simulations become an effective tool for studying the system in detail. They can also provide valuable insight into the kind of material parameters that need to be synthesized for the purpose of application in devices. Here we present the results of Monte Carlo simulations based on a random anisotropy model for a single domain magnetic array, with anisotropy and  interparticle interactions like long range dipolar and short range  exchange. The simulations of the ZFC -  FC magnetization  are done for different values of the exchange parameter and the results appear to be in good agreement with the experimental results.

In Section 2  we discuss the experimental techniques, in Section 3 we discuss the results of the  characterization and magnetic  studies performed on the samples.   In Section 4 we discuss the method and results of simulation   and  finally, in Section 5  we present our conclusions.

\section{Experimental}

Maghemite  nanoparticles were prepared by the sol - gel method using ferric nitrate ($Fe (NO_3)_3.9 H_2 0$) as  precursor  and 2 - methoxy ethanol as solvent  \cite{p1,p2,k1}. The powder obtained by this method was suspended in distilled water and stirred using a magnetic stirrer for five hours. The above suspension was placed in a sonicator for 10 minutes. This resulted in a nonuniform suspension. This suspension was  centrifuged at 5000 rpm for 20 minutes  and further decanted.  The remaining powder was heated  to remove water and was used for coating with polymer.

1 gm of the dried powder  was dispersed in a mixture of ethanol  and deionized water in the ratio 7 : 10. About $0.2$ ml of  pyrrole monomer (Aldrich reagent grade, vacuum distilled, weight per ml of pyrrole is $\sim 0.969$ gm )  was added to the above mixture. This was followed by the addition of $4$ cc of $0.4 \%$ PVA  solution. The PVA only acts as a binding agent and has no effect on the polymerisation \cite{k3}.
A uniform mixture  was obtained by placing  the above mixture  in an ultrasonic bath. This system was further diluted to $10$ cc with deionised water. The whole mixture was then sealed tightly in a reaction vessel and heated at $85 ^0$ C for $10$ hours while being stirred continuously  with a magnetic stirrer. The resulting solution was filtered and the powder obtained was washed thoroughly with distilled water. This method was followed  for different pyrrole concentrations. Initial concentrations of  iron oxide :  pyrrole (monomer) were maintained in the ratios of $1 :0.2$, $1 : 0.8$, $1 : 1.2$ , $1 : 1.6$  and $1 : 2.0$.

The structural properties of the  coated particles  were investigated by  XRD, using  a Rigaku Rotaflex Diffractometer with a Cu - $K \alpha$ radiation ($\lambda = 1.54056 $ A). TEM was used to study the shape, size and morphology of the particle and was performed using a  JEOL - JEM 2000 EX. The presence of polypyrrole was confirmed using  FTIR, which was recorded using a Perkin Elmer FT-IR spectrometer, Spectrum 2000 model. A  Rigaku PTC - 10 A thermobalance was used for investigating the degradation of the polymers.  The magnetic measurements were performed in the temperature range $77$ K to $300$ K using a Lakeshore Cryotronics INC model 7000 AC Susceptometer. The measurements were performed at a field of  $800$ Amp/m  and at a frequency of $16$ Hz.  

\section{Results and Discussions}
Our aim has been to obtain iron oxide polypyrrole nanocomposites with less agglomeration and retain the magnetic phase. In the above method it was observed that all the prepared samples were magnetic. The iron oxide powders obtained by the sol - gel process was light brown in color. With increasing concentration of pyrrole the color of the powders was found to change to dark brown and for concentrations higher than $1: 1.6  $ they were found to be black, showing the presence of more polymer.  As mentioned earlier PVA has been used only as a binder in our preparation and does not affect the polymerisation process. After confirming the degradation temperature of PVA by TGA - DTA measurements, all the samples were annealed at $250^0$ C to remove the excess PVA. The conductivity was measured  and  found to be very high ($\sim $ M ohms), since no dopants were present in any of the composites.

The XRD studies performed on some of the nanocomposites is shown in Fig. 1. It is observed that all the annealed samples, annealed at 250 $^0$ C, retain the cubic  $\gfe2$ phase. The lattice constant calculated from the diffraction pattern is  $8.346  \pm .0001$ A and $c=25.02 \pm .0001$ A and  agrees well with reported  values \cite{new1}. The grain size as estimated from the Scherrer's  formula is $\sim 20$ nm. 

In order to confirm the presence of polypyrrole the FTIR measurements were   performed on the unannealed sample  using KBr pellets.  Fig 2 shows the FTIR for one of the nanocomposites  namely $ 1:1.6$. The  peaks at $1096$, $1620$ and  $3401$  $ cm^{-1}$ are identified as  the $C=C$, $N-H$ and $C-N$ bonds,  respectively of the polymer backbone. The peaks appearing at $2343$ and $2923$  $ cm^{-1}$ belong to the $C - H$ and $-OH$ bonds  of PVA. The peaks between  $400$ to $700$  $ cm^{-1}$ corresponds to the $Fe - O$ bonds.

Fig 3(a) shows the TEM of the $\gfe2$ obtained from the sol - gel process which shows agglomerated  iron oxide particles. Figure 3(b) shows the TEM for the smaller sized particles obtained after stirring, sonicating and centrifuging these particles. Although the grain size is reduced there is still agglomeration present.  The TEM performed on one of the nanocomposite  samples ($1: 1.6$) is shown in Fig 3(c) which is seen to have much less  agglomeration.

The TGA - DTA measurements were performed on several samples of varying concentrations of pyrrole in order to estimate the degradation temperatures of PVA and PPY.  Fig 4(a) shows the DTA  for  one of the unannealed  nanocomposites ($1: 1.6$). It  shows  exothermic peaks  at $210^0$, $300^0$ and $350^0$ C which corresponds to the degradation of PVA, degradation of pyrrole and phase change to $\alpha - Fe_2 O_3$  respectively. The  weight loss obtained from the TGA measurements indicates the amount of polymer present in the composites and this is indicated in detail in Figure 4 (b). The weight  is found to increase with increasing monomer concentration saturating beyond a concentration of $1:1.6$  (Figure 4(b) inset). 

Since we have carried out the oxidation of pyrrole only in the presence of air, it has been observed that even with an increasing monomer concentration the polymerization is very less. This is confirmed by the measurements above. The IR however shows the presence of PPy and hence it is likely that PPy grows around iron oxide core. This is markedly different from our observations on these composites when they were prepared by the simultaneous gelation - polymerization process which showed chain structure \cite{k1}.

In Fig 5(a) and (b) we present the ac susceptibility results for two different concentrations namely $1:0.2$  and $1:1.6$ . For the low concentration sample the susceptibility increases with  temperature till $\sim 300$ K. However for the higher concentration the $\chi ^\prime$ is found to increase and at $\sim 180$ K it starts decreasing, showing a clear transition to the superparamagnetic phase which is not evident in the sample with low concentration. The blocking temperature for the higher concentration as deduced from the susceptibility measurement is $180 $ K. The blocking temperature is not sharply defined, which is expected since we have a distribution in the particle sizes in the sample. In the low concentration samples there is a large amount of agglomeration, because of which  the exchange interactions are  present. This results in a higher blocking temperature \cite{d2,d3}. Such behavior is also observed in rare earth clusters where even for very small particles ($\sim 2 $ nm) the blocking temperature observed is as  high as $400$ K  \cite{r10}. The effect of coating of the $\gfe2$ particles with the polymer has resulted in smaller clusters with lesser interactions and hence lower blocking temperatures.

We have carried out Monte Carlo simulations on an interacting, random anisotropy model to study the effect of interactions on the blocking temperature of single domain magnetic particles. 

\section{ Simulation}
\label{sec4}
The model  Hamiltonian for a system of interacting single domain magnetic particles, each having a magnetic moment vector  $\vec \mu_i  $  can be  written as,\cite{r19},
\begin{eqnarray}
H&=&- K  \sum_i V_i \left( {\vec \mu_i . \vec n_i  \over |\vec \mu_i|} \right)^2 - \sum _{<i\neq j>}J _{ij}\vec \mu_i . \vec \mu_j\nonumber \\
&-& \mu _0  \sum_{<i\neq j>}
{3 (\vec \mu_i.\vec e_{ij})( \vec \mu_j.\vec e_{ij})- \vec \mu_i. \vec \mu _j \over r_{ij}^3} \nonumber \\
 &-& \mu_0  \sum _i \vec  H . \vec \mu_i \nonumber \\
\end{eqnarray}
\noindent The first term in Eq. 1 represents the anisotropy energy of the $i$ th  magnetic particle which has  $K$ as the anisotropy constant and $V_i$ as its volume. Its magnetic moment vector is $\vec \mu_i$ and the direction of the easy axis of magnetization of the particle is represented by the unit vector $\vec n_i$. The second term is the exchange interaction  energy between the  different particles in the array and  $J_{ij}$ is the strength of the  ferromagnetic exchange interaction between two  particles with localized magnetic moment vectors   $\vec \mu_i$ and $\vec \mu_j$ respectively. The third term  is the dipolar interaction between these particles, with $r_{ij} $ as the distance between the i$th$ and j$th$ particles and  $\vec e_{ij} $  the unit vector pointing along $\vec{r_{ij}}$. The last term is the energy of the particles due to an externally applied magnetic field $\vec H$.  
For the purpose of simulation we assume that  the magnetic moment vector  for a single particle has a temperature independent  magnitude and  $ \vec \mu_i = V_i M_S\vec  \sigma_i$ where  $M_S$ is the saturation magnetization and $\vec \sigma_i $ is the unit vector along the direction of magnetization. We also assume that the exchange interaction has a site independent constant value $J_{eff}= J $.  The method of preparation and the experimental observations indicate that  magnetic nanoparticles in the sample are (i) not all of the same size and (ii) positioned randomly in the sample.
Accordingly we  work with a basic simulation cell which is  a cube of size $L^3$ in which  $N= 64$  single domain magnetic particles   are  randomly distributed. The volumes of the particles are picked from a normal distribution $P(V)dV ={1\over (2\pi t^2)^{1/2} }\exp \left(- {(V-V_0) ^2 \over 2 t^2}\right) $ where $V_0$ represents the mean volume  of the particles which is taken to be equivalent to a sphere of diameter  $15$ nm and $t$ is the width of the distribution which  is taken to be $.1$. The directions of the easy axis of magnetization of the particle are also picked randomly The dipolar interaction energy is calculated by summing over periodic repeats of the basic simulation cell by the method of Lekner summation \cite{r21,r22}.

The simulation  of  Field Cooled - Zero Field Cooled (FC-ZFC) magnetization is carried out by the Monte Carlo method using the standard Metropolis algorithm \cite{r10,r23,new10}. The value of $M_S =4 \times 10^ 5 $ Amp/m  for $\gamma -  Fe_2 O_3$ is  taken from literature. The magnetic field is kept fixed at $.01$ Tesla 
To fit to the experimentally observed values of the blocking temperatures in these systems we find that the anisotropy constant should be higher the reported values for the pure $\gfe2$ system which is $\sim .045  \times 10 ^5$ $J/m^3$. We find that  appropriate results are obtained for a much higher value of the anisotropy constant , viz.  $K = .1 \times 10^5 $  $J/m^3$. This is keeping with the general trend seen that the  values of the anisotropy constant observed  for  superparamagnetic iron oxide  nanoparticles  in a nanocomposite are  usually quite high  and for small particles of $\gfe2$ and $Fe_3 
0_4$  can be an order of magnitude larger than the values for bulk, in fact   high as  $2 - 4 \times 10 ^5 $ $J/m^3$ \cite{a1,a2}.
The value of  $J_{eff}$ is input as a parameter in the simulation.  
 The purpose of the simulation is to estimate the blocking temperature  from the FC- ZFC magnetization curves which are shown in Fig. 6, where we plot the variation of the scaled magnetization $M/M_S$ with the temperature. 

The relevant interactions in an assembly of single domain particles are the dipolar interactions and in this case, since there is clustering of the particles, exchange interactions. We find that the value  of the blocking temperature is decided  mainly by the anisotropy and  the strength of the exchange interactions and not  much by the dipolar interactions. The effect of  dipolar interactions is more pronounced in the reduction of the  magnetization.In curve (a) of Figure 6 we plot the FC - ZFC magnetization for a system in which the dipolar and exchange  interactions are negligible (that is for a very dilute system of particles). We find that the blocking temperature $T_B$  for this configuration is $\sim 150$ K. In curve(b) the dipolar interactions  between the single domain particles are included in the simulation  but the  exchange interactions are neglected, the  other parameters of the simulation remaining same as for curve(a). We find that (i) the magnitude of the magnetization is reduced and  (ii) there is hardly any  change in the value of $T_B$ as compared to (a). In curves (c) and (d) we have plotted the FC - ZFC magnetization for the same simulation system for  two different strengths of the exchange interaction  
 $J$ which are  $J=.05 E_A$ and $J= .1 E_A$. $E_A$ is the anisotropy energy of a spherical particle of radius $ R_o = 7.5$ nm which is $K (4/3 ) \pi R_o^3$.  The curves show  some distinct features namely (i) the magnetization is enhanced compared to curve (b)  which is expected because of the cooperative exchange interaction between the particles (ii) the blocking temperature is much higher, in range of $190$ to $220$ K (iii) the  curves show a peaked structure  and (iv) the FC curves show a double peak structure which is more pronounced in (d) and the same is reflected in the corresponding heating curve. The values of the $T_B$ obtained from our simulation for (c) and (d)  match well with experimental values (fig. 5). The enhancement of blocking temperatures and  as well as the peak in the magnetization curves are known to be the effect of the  interactions \cite{d1,newp1}. In particlular, we feel that the double peaked structure  below the blocking temperature which is seen only in the system with interactions and is more pronounced in the system with stronger interactions  is a manifestation of  competing interactions like dipolar and exchange in a system which further has strong disorder. It could also indicate  some collective behaviour or memory-like effects, which however can be confirmed through further simulations which are now under progress. Similar  behavior has also been observed experimentally recently \cite{new11,new12}. We are in the process of conducting more detailed simulations to study the low temperature behaviour of these systems in more detail.  

\section{Conclusions}
 Polypyrrole coated iron oxide nanoparticles of varying monomer concentration were  prepared and characterized using X-Ray, IR and TEM. The FTIR shows the presence of the polymer. The DTA - TGA analysis was used to determine the degradation temperature of polypyrrole which was found to be $\sim 300 ^0$ C. Subsequently the samples were annealed at $250^0$ C for further investigations. The X- Ray studies confirmed that the original  $\gfe2$ phase was retained for all the  composites prepared unlike in the case of simultaneous  gelation - polymerization methods. TEM investigations clearly indicates smaller cluster sizes  for higher concentrations of pyrrole. The ac susceptibility measurements performed on the uncoated and low pyrrole concentration sample did not show any blocking in the range of temperature $77$ to $300$ K. Samples with higher pyrrole concentration are superparamagnetic and  show blocking at fairly high temperatures $\sim 180$ K.  We feel that the higher blocking temperatures are a result of (i) larger anisotropy which is known in nanocomposites and (ii) cluster effects  which lead to strong interparticle interactions. These results are supported by  our simulation results which give blocking temperatures in the the range of $190$ to $220$ K depending on the strength of interaction. Hysteresis measurements are in progress to investigate the coercivity as a function of pyrrole concentration. By further refining the method of preparation of the composites as well as the matrix materials which are known to affect the anisotropy,  we hope to obtain well separated smaller domains which would be  of use in magnetic memory devices and refrigeration. Since the conducting polymers are organic semiconductors, dispersing magnetic nanoparticles in these would make them an interesting material for spintronic devices. 

{\bf Acknowledgements} 
One of the authors R S acknowledges the CSIR for providing a fellowship. The authors wish to thank Mr. B. V.  Kumaraswamy, National Physical Laboratory, New Delhi for ac susceptibility measurements.
We wish to acknowledge the Department of Science and Technology (DST) for their
financial assistance throught their Project (SR/S5/NM-52/2002)  from their Nanoscience and Technology Initiative programme. 
We also wish to thank  Dr. N. C. Mehra and Dr. S. K. Shukla, University Science Instrumentation Center, University of Delhi for their help in carrying out the TEM and X- Ray studies.

\newpage

\begin{figure}
\caption{ XRD patterns of (a) $\gfe2$  and  polypyrrole coated $\gfe2$ of concentration  (b) $1:0.8$ (c)  $1:1.6$  and (d)  $1:2.0$ , all annealed at $250 ^0$ C. }
\end{figure}

\begin{figure}

\caption{ FTIR spectra of unannealed  polypyrrole coated $\gfe2$  of concentration $1:1.6$. }
\end{figure}

\begin{figure}
\caption{ (a) TEM of unannealed $\gfe2$ as obtained by sol -gel process  (b) TEM   of unannealed  $\gfe2$  after magnetic breaking  and (c)TEM of polypyrrole -  $\gfe2$ nanocomposites  of ratio $1:1.6$. }
\end{figure}

\begin{figure}
\caption{(a) DTA of  unannealed polypyrrole -  $\gfe2$ nanocomposite  of ratio $1:1.2$ (b)TGA of (1)pure iron oxide and   $\gfe2$ -polyrrole  nanocomposites of following  concentrations (2)  1.0:0.2 (3) 1.0:0.4 (4) 1:1.2 and (5) 1:2.0,   inset shows the weight loss percentage with monomer volume in ml }
\end{figure}

\begin{figure}
\caption{Variation of ac susceptibility  with temperature for   polypyrrole coated $\gfe2$ annealed at $250^0$ C of concentration  (a) $1:0.2$ and (b) $1:1.6$. }
\end{figure}

\begin{figure}
\caption{Variation of normalized  magnetisation $M/M_S $ with temperature $T$ in  K  for a system with  (a)no interactions(dipolar or exchange)    (b)only dipolar interactions (c) dipolar interaction and exchange interaction of strength  $J= .05 E_A$ and (d) dipolar interaction and exchange interaction of strength $J=.1 E_A$. }
\end{figure}

\end{document}